# Interplay of Spin, Lattice, and Charge Degrees of Freedom in $Ca_3Mn_2O_7$

Pooja Sahlot[1], A.K. Sinha[2], and A.M. Awasthi[1]*

[1]*UGC-DAE Consortium for Scientific Research, University Campus, Khandwa Road, Indore- 452 001, India*

[2]*Indus Synchrotrons Utilization Division, Raja Ramanna Centre for Advanced Technology, Indore- 452 013, India*

*amawasthi@csr.res.in

**Abstract:** From low-temperature Synchrotron X-ray diffraction, a precise thermal characterization of octahedral distortions in single phase Ruddlesden-Popper $Ca_3Mn_2O_7$ is performed. Highly sensitive close-steps temperature dependences of Mn-O-Mn bond angles connecting MnO6 octahedra clearly reveal signature of the spin-ordering in the system. Spin-lattice coupling is thus established via the structural distortions responsible for evolution of the magnetic state. Further, temperature anomalies observed here in volume and polarization-measure of the unit cell highlight the interplay between spin, lattice and charge degrees of freedom. Dipole-relaxation characteristics examined under applied magnetic field consistently corroborate the concurrent magnetic and structural changes, in terms of genuine and intrinsic magneto-dielectricity.

**Keywords:** Ruddlesden-Popper Layered Ceramics; Dielectric Relaxation; Spin-lattice Coupling.

## INTRODUCTION

Octahedral rotations greatly affect magnetic properties, and systems where external field can couple to these rotations, cross-coupling between magnetic and electric properties can be identified. Here, we have studied a layered perovskite, in which octahedral rotations can induce polarization, which directly sees applied electric field. Ruddlesden-Popper (R-P) compounds ($A_{n+1}B_nO_{3n+1}$) recognized as layered systems are getting much attention because of their evolved complex magnetic [1,2] and electrical properties [3]. These perovskites-related systems possess structural-distortion coupled range of magnetic phases. Compared to the 3D perovskites, these systems are structured as oxygen-octahedron, consisting $B^{4+}$ ions in single- or multiple-connected perovskite blocks, sandwiched in between $A^{2+}$ cation-oxygen layers [4,5]. Using the first-principles calculations [5], $n = 2$ R-P compounds have been found to support hybrid improper ferroelectricity (HIF), evident experimentally [6,7]. HIF mechanism describes ferroelectricity with a combination of two inequivalent symmetric octahedral-distortion patterns [8]. Reports on $n = 2$ RP compound $Ca_3Mn_2O_7$, possessing tetragonal phase (I4/mmm) at high temperatures, have established introduction of two symmetrically different octahedral distortions; which result in the transition to orthorhombic broken-centrosymmetry phase (Cmc21) at lower temperatures [9]. Among these distortions, oxygen octahedral-rotation has been found to enhance magneto-electricity in the system. Along with it, oxygen octahedral-tilt is introduced, which supports canted spin moments in the system at low temperatures, explained with Dzyaloshinskii's criteria [8]. Depending upon the metal-oxygen-metal bond angle, unfilled $Mn^{4+}$ d-orbitals in oxygen-octahedron environment adopt (anti)ferromagnetism [10]. Furthermore, Mn-O-Mn interatomic distance and octahedral-distortion were explored with Th-doping at Ca-site and more decisive study on structure has been motivated [11].

Magnetic state predicted in $Ca_3Mn_2O_7$ using first principles calculations [8] was confirmed by Lobanov et. al., via its temperature dependent neutron diffraction study, establishing G-type antiferromagnetic (AFM) state [12]. Symmetry conditions further permitted weak ferromagnetism (WFM) on lowering the temperature. These initial studies open scope for more precise thermal evolution of structure, which can be analyzed to witness the coupling between structural, magnetic and electrical properties of the system. Recently in $Ca_3Mn_2O_7$ we have reported antiferromagnetic transition with Néel temperature 123K, along with the emergence of weak ferromagnetism (WFM) below 110K [13,14]. The WFM is traced to the canted spin-clusters in the AFM-matrix, rather than to a uniform spin-canting. This is consistent with the observed exchange-bias (EB) effect, characterized in detail for the system. The indirect spin-exchange interactions are profoundly affected by the change in transition metal-oxygen-transition metal bond-length and -angle, which get altered by the octahedral-distortions [15,16]. To examine the inter-coupling of structural distortions with magnetic and electrical properties of the system, it is essential to investigate the changes in the structural environment of $Mn^{4+}$ ions, versus thermal cooling of the system. The system possessing broad structural transition [7] is expected to feature short range polar correlations, promoted to long range ordering at low temperatures, with well-established orthorhombic phase. But the system is observed to have substantial polarization only below the magnetic ordering. Semi-conducting nature at high temperatures is expected to be the cause of elusive polarization. At low temperatures, cross-coupling between structural, magnetic, and electric properties is studied. Here, we intend to analyze the crystallographic structure of $Ca_3Mn_2O_7$ system with temperature, investigated using Synchrotron powder X-ray diffraction (SPXRD). Further, to analyze the coupling between magnetic and electrical properties, electrical conductivity measurements under the application of 6Tesla magnetic field have been conducted. In this communication, we notice that the strong inter-atomic spin-lattice coupling, deduced from the SXRD results, is not optimally translated into macroscopic magneto-electric coupling. This circumstance is essentially attributed to anti-ferroelectric configurations of polarized layers within the unit cell. Studies performed with non-magnetic B-site cation substitution could increase the polarization, albeit at the cost of magnetization. To enhance magnetically coupled polarization, we suggest part-substitution of AO-layers by A'O layers.

## EXPERIMENTAL DETAILS

Single phase polycrystalline $Ca_3Mn_2O_7$ compound has been prepared with high purity (99.5%) $CaCO_3$ and $MnCO_3$ ingredients using solid-state synthesis, reported in detail in [13]. Using a Quantum Design SQUID-VSM, zero-field cooled (ZFC) and field cooled (FC) $M(T)$ were measured. For temperature-dependent structural analysis, Synchrotron powder X-ray diffraction (SPXD) has been performed and analyzed in the temperature range of 37K-340K. SPXD has been performed on area detector (Image Plate Mar-345) using wavelength 0.79866Å. Rietveld analysis was performed using Fullprof program. Low temperature ac-permittivity measurements were performed using Novocontrol Alpha-A Broadband Impedance Analyzer, under zero and 6T field, using Oxford Nanosystems Integra magnet-cryostat.

## RESULTS and DISCUSSION

### Magnetic Properties

In a report [14] on $Ca_3Mn_2O_7$ depicting dc magnetic susceptibility ($\chi = M/H$) measurement (under 100 Oe field), long range AFM state below 123K with WFM below 110K has been established. AFM state's non-robustness is illustrated by the high frustration parameter $f = |\vartheta_{C-W}/T_N| = 7.2$. In fig.1 we show our FC $\chi$-$T$ curve, along with the ZFC $\chi$-$T$, measured under high (7T) magnetic field. A hump around ~150K, corresponding to short range AFM correlations with a small peak at ~107K ($\equiv T_N|_{7T}$) is seen, with no ZFC-FC bifurcation. Here, for 7T-cooled $M(T)$, the AFM transition temperature is down-shifted and merged with the emergence of WFM (split $M$-$H$, fig.1 inset) at 108K. Overlap of the ZFC-FC curves under 7T field indicates that the corresponding Zeeman energy surpasses the energy barrier (corresponding to FC-ZFC split). As the temperature is increased, thermal fluctuations support rotation of some of the spins along the direction of the applied magnetic field, leading to the reduction in split seen with temperature in $\chi(T)$ under 100 Oe field, as reported earlier [14]. Similarly, here a suppression of splitting is effected with the application of high magnetic field of 7 Tesla. Upturn seen below 90K is due to the paramagnetic impurity.

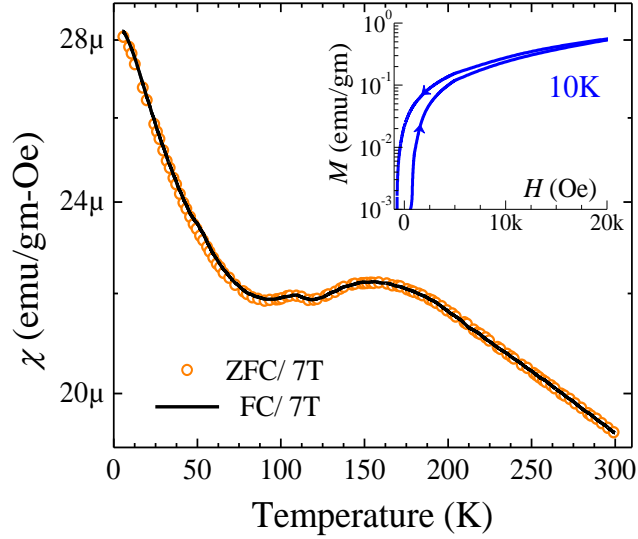

Figure 1. ZFC/FC-DC magnetic susceptibility $\chi(T)$ under 7T field. Inset: *M-H* loop (first quadrant, on log-lin scale) at 10K.

## Structural Characterization

The X-ray diffraction profile for the system is refined under Cmc2$_1$ orthorhombic symmetry from 300K down to 37K. The model fit for SPXD pattern at 300K is shown in fig.2(a), with acceptable value of goodness of fit ($\chi^2$ =1.11). Figure.3(a) shows schematic for the crystal structure obtained from the fitting. Rietveld fit performed gives the thermal evolution of unit-cell parameters in the system. The lattice parameters *a*, *b*, and *c* decrease monotonically below 300K, whereupon cooling by 130K, the slopes of their decrease substantially reduces, as shown in fig.2(b). Cell volume too shows shrinkage of the unit cell upon cooling. The (linear) decrease of volume with temperature reduces below 123K. This may enhance the FE-stability to prevail over the antiferrodistortive features present in the system, which has been well established in perovskite structures possessing both these tendencies [17]. Evidence of this has been observed from thermal evolution of the dielectric relaxations, explaining the formation of polar nano-regions (PNRs) below 123K, in a previous report [18].

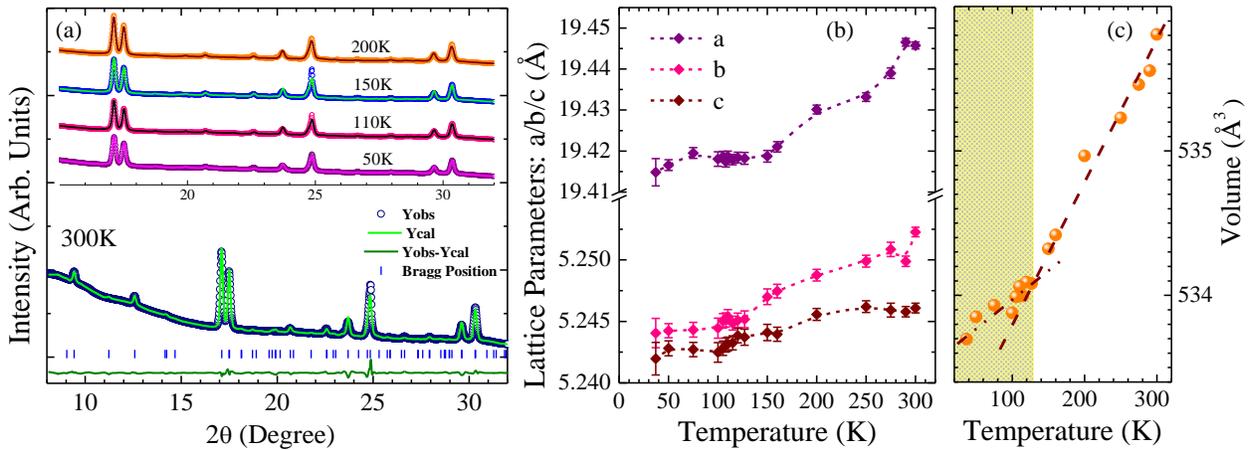

Figure 2. (a) XRD plot at selected temperatures, with XRD fit at 300K. (b) Temperature dependence of lattice parameters obtained from Rietveld fitting. (c) Temperature dependence of unit cell volume obtained from Rietveld fitting.

The system has been reported to undergo G-type AFM transition at $T_N$ =123K, with a coexistent WFM phase emergent upon further cooling [13]. Structural distortions related to the oxygen octahedra are expected to play a crucial role in changing the electrical and magnetic properties of the system. Mn-O-Mn bond angles demonstrating the effect of octahedral distortions on the inter-octahedral correlations have been analyzed with temperature, and are shown in fig.3(b). Schematic for crystal structure of $Ca_3Mn_2O_7$ in fig.3(a) shows the connectivity of MnO6 octahedron in the *bc*-plane via O2 oxygen atom and along the *a*-axis via O1 oxygen atom. Here, profound changes in the Mn-O2-Mn and Mn-O1-Mn bond angles upon cooling reflect remarkable octahedral modulation. Increment in Mn-O2-Mn (*bc*-plane) bond angle upon cooling from ~143° (at 300K) to ~146° (at 37K) supports AFM ordering in the *bc*-plane, in accordance with that observed for the system [13, 14]. Mn-O1-Mn (along the *a*-axis) bond angle initially increases upon cooling from ~173° (at 300K) to ~176° (at 120K), supporting AFM ordering along the *a*-axis. While on further cooling, Mn-O1-Mn bond angle decreases to ~170°. The effective reduction in the metal-oxygen-metal bond angles, formed from the interconnected oxygen octahedra in the octahedron layers, signals inclination towards ferromagnetic (FM) feature along the *a*-axis. Here, these experimental observations highlight the $MnO_6$-octahedra changes; profoundly affecting the magnetic state in the orthorhombic phase of the system.

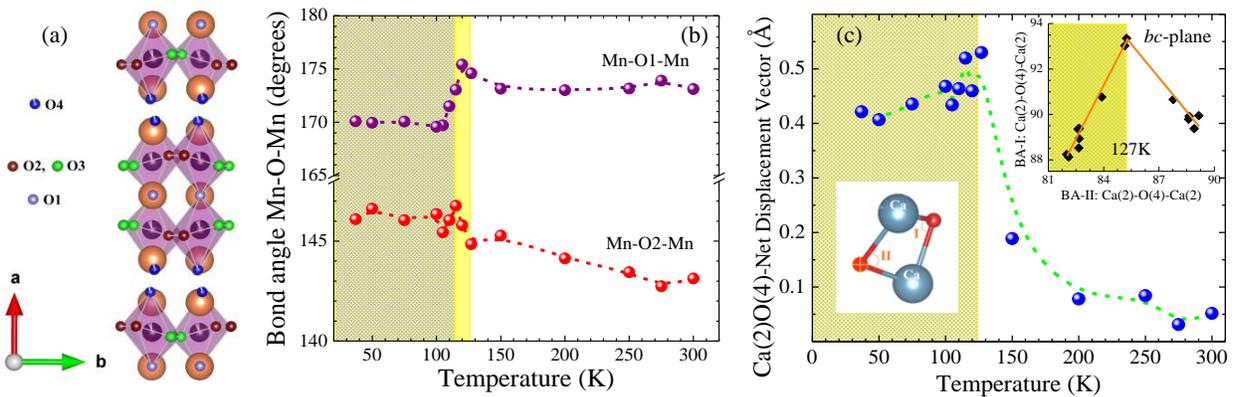

Figure 3. (a) Schematic of orthorhombic crystallographic structure of $Ca_3Mn_2O_7$ depicting MnO6 octahedra in *ab*-plane. (b) Temperature dependence of bond angle Mn-O1-Mn and Mn-O2-Mn connecting MnO6 octahedra along the *a*-axis and the *bc*-plane respectively. (c) Temperature dependence of net Ca(2)-O(4) displacement vector in the *bc*-plane, as a measure of the individual-layer polarization. Inset: Opposite regressions shown above and below ~$T_N$, between the bond angles BA-I and BA-II, formed at the O(4)-oxygen of the Ca(2)-O(4) layers in the *bc*-plane.

To analyse the modulation in dielectric polarization for the system, the net displacement-vector for Ca(2)O(4) layer in the *bc*-plane has been evaluated. In Ruddlesdon-Popper materials, polarization has been stated as proportional to the displacement of A-cation in AO-layer at the interface [5]. Figure.3(c) shows this displacement vector for Ca(2)O(4) layer versus temperature, with abrupt enhancement across 127K. Plot of the bond angles Ca-O-Ca in the *bc*-plane, shown in fig.3(c)-inset again emphasizes the behavioral change in electrical properties with magnetic ordering. Thus, magnetic ordering in the system conjointly influences its electrical properties via substructural changes, as has been observed in dielectric studies of the system [18].

Figure.4 presents SPXRD patterns at and above the room temperature, up to 340K. Above room-temperature diffraction patterns exhibit weak reflections, indexed as (0 0 10) for I-centered tetragonal symmetry (I4/mmm). This indicates the presence of mixed phase above room temperature. Peaks for both the mixed phases are clearly resolved and the refinement resulted in the structure as an amalgamation of the orthorhombic Cmc21 (majority) phase with weight fraction of 98.48% and tetragonal I4/mmm (minority) phase with weight fraction of 1.52%. The amalgamation of orthorhombic and tetragonal phases was observed even at 340K, indicating the presence of coexistent crystal structures at higher temperatures, which is consistent with the previous reports. [19].

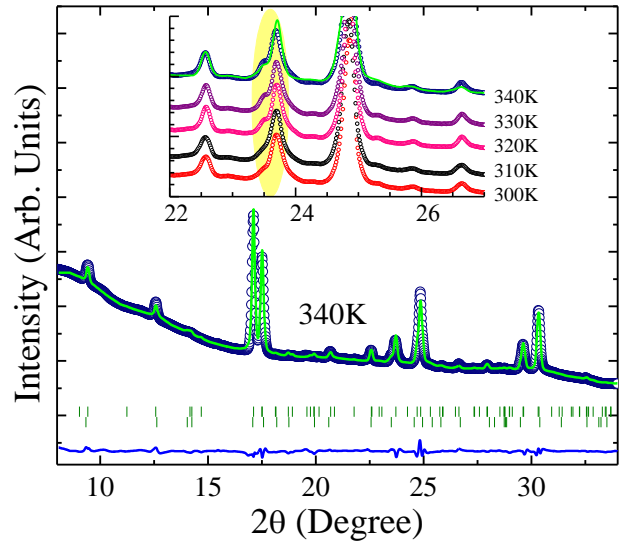

Figure 4. Fitted SPXRD pattern for $Ca_3Mn_2O_7$ at 340K; Inset: SXRD patterns obtained at 300K and above, depicting the introduction of the tetragonal phase.

**Magneto-electricity**

To explore the coupling between electrical and magnetic properties in the system, low temperature dielectric measurements were performed. Figure 5.(a), (b) show isothermal spectra of real part of dielectric permittivity and loss-tangent, respectively. Relaxation peaks in loss tangent ($\tan\delta = \varepsilon''/\varepsilon'$) are observed. Insets in fig. 5(b) show $\tan\delta$ isotherms under zero- and 6T applied magnetic field. Increase in the peak frequency of $\tan\delta_T(f)$ isotherms with rise in temperature depicts thermally activated character. Upshift of these isotherms to higher frequency-side under the applied *H*-field indicates speedup of relaxations, due to reduced barrier-activation energy, discussed later.

Dielectric relaxations are studied comprehensively in the imaginary dielectric modulus *M''*, shown in fig.6(a) and fitted with empirical Havriliak-Negami (H-N) function (WinFit, NovoControl). The model expresses the relaxation function *M\** in the following form [20];

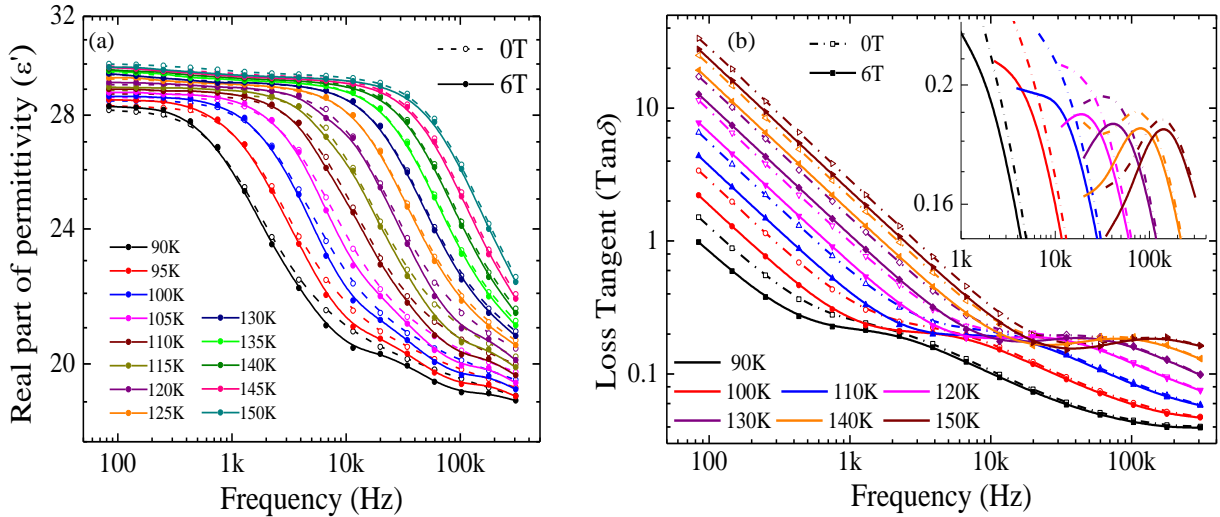

Figure 5.(a) Frequency dependent $\varepsilon'$ spectra at various temperatures (90K to 150K) over 100Hz to 300kHz under zero- and 6T magnetic field. (b) Frequency dependent loss-tangent spectra over 100Hz to 300kHz under zero- and 6T magnetic field. Inset shows zoomed-in view at the mentioned temperatures, depicting relaxations at 0T and 6T-field.

$$\frac{M^* - M_\infty}{M_0 - M_\infty} = \frac{1}{(1 + (i\omega\tau)^\alpha)^\beta}$$

Here, $\omega$ (=$2\pi f$; $f$ = frequency of the applied electric-field) with $\omega_{max}\tau \approx 1$; $\omega_{max}$ denoting the maximum-relaxation frequency, marking the transit from the high-frequency/unrelaxed modulus ($M_\infty$) to the low-frequency/relaxed modulus level ($M_0$). Exponent $\alpha$ measures the coupling between multiple dipoles ($\alpha < 1$ signifies extra-Lorentzian spectral peak-width) and $\beta$ is the asymmetry parameter of the relaxation peak. H-N fit incorporates the Jonscher power-law frequency-dependence of ac-conductivity, as explained later.

No direct signal of long range electrical ordering is observed from the dielectric modulus function either. Imaginary part of the modulus shows two relaxation processes for the system; at low-frequencies ($R_1$) and at high-frequencies ($R_2$), fig.6(a). H-N fits performed over 90-140K temperature window for $M''(f)$ yield non-trivial ($\alpha, \beta < 1$) exponents' values for both $R_1$- and $R_2$-relaxations, revealing correlations among the dipoles.

Conductivity parameters obtained from the H-N fit are also studied. In such Mott insulators, conductivity is explained using Pollak's theories [21, 22]; well described by its dc- and ac- components, which correspond respectively to pinned and free dipole-relaxations [23, 24]. Jonscher power law $\sigma_{ac} = A\omega^n$ [25] provides the empirical power-law exponent $n$, whose temperature dependence characterizing the short-timescale current response, is used to divulge the particulate conduction mechanism pursued by the system. "Universal law of dielectrics" [26] well describes the total electrical conductivity for the system:

$$\sigma(\omega) = \sigma_{dc} + A\omega^n = \sigma_{dc}[1+(\omega/\omega_h)^n]$$

Here, $\omega_h$ defines the hopping frequency at the crossover from dc- to ac-conduction ($\sigma(\omega_h) = 2\sigma_{dc}$), which shifts to higher values with increase in temperature. For the correlated electron system in $Ca_3Mn_2O_7$, multiple hopping mechanism [27,28] between the charge-carrier sites explains the ac-conductivity.

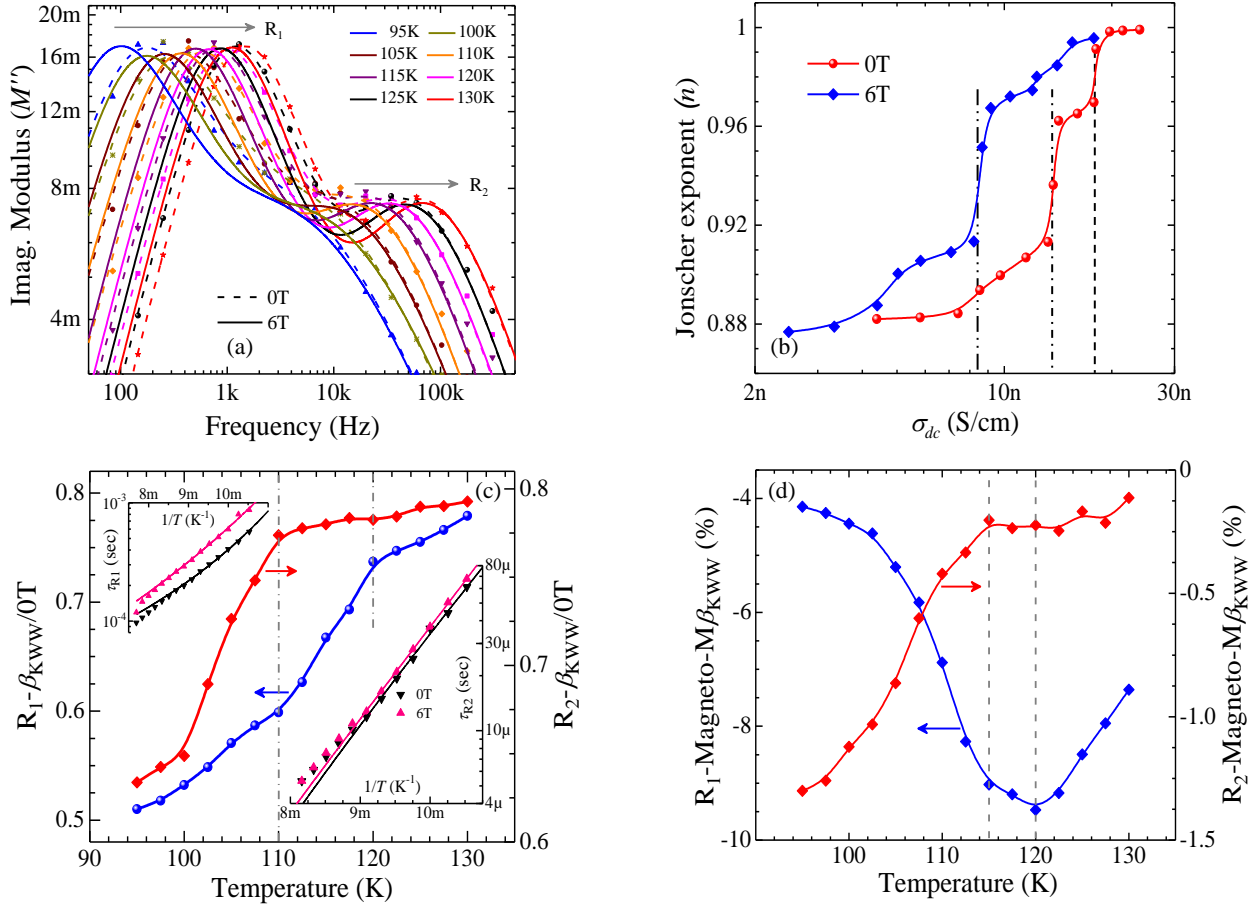

Figure.6.(a) Frequency spectra for imaginary part of dielectric modulus $M''_T(f)$ with Havriliak-Negami (H-N) fit at the mentioned temperatures. (b) Jonscher power-law exponent '$n$' vs. $\sigma_{dc}$ for measurements under zero- and 6T-field. (c) Temperature dependence of stretched-exponent parameter $\beta_{KWW}$ corresponding to the relaxations $R_1$ (left-axis) and $R_2$ (right-axis). Insets: $\tau$ vs. $1/T$ (log-lin) plots of characteristic timescales for $R_1$- (top-left) and $R_2$- (bottom-right) relaxations. (d) Temperature dependence of magneto-beta $M\beta_{KWW}$(%) for $R_1$-relaxations (left-axis) and $R_2$-relaxations (right-axis).

In the present context, conductivity-signatures of magnetic evolution in the system are best illustrated by contrasting its dc/pinned-dipole and ac/free-dipole relaxation-characteristics. For this purpose, we show in fig.6(b) the Jonscher-exponent $n(T)$ vs. $\sigma_{dc}(T)$, both determined from H-N fits on the $M''_T(\omega)$ isotherms at zero and 6T $H$-fields. Both $n$ and $\sigma_{dc}$ decrease upon cooling; though $n(\sigma_{dc})|_{H=0}$ is marked by a sharp step across the AFM transition. Upon further cooling, another sharp step follows-- concurrent with the emergence of the WFM-phase within the AFM matrix. Nearly-vertical steepness of these steps evidences that magneto-electricity influences the short-timescale/ac electrical transport exclusively. The two successive steps observed under $H=0$ merge to a single step across $T^* \sim 115K$ at 6 Tesla field. This is consistent with 'merging' of suppressed-$T_N$ and enhanced-$T_{WFM}$ under high $H$-fields; as also observed in $Gd_5Sb_3$ [29].

Note that 'ironically-negative' magneto-$\sigma_{dc}$ found here but confirms the isolation of 'less-insulating' WFM nano-regions-- which rather act as scatterers of dc charge-carriers; vis-à-vis as generators of continuous 'metallic' pathways, in otherwise uniformly-canted antiferromagnets. Both in the paramagnetic ($> T_N$) and the well-established AFM(+WFM) ($< T_W$) phases, absence of magneto-electric effect renders the 0T/6T-adjoined $n(\sigma_{dc})$ co-regressions featureless. Contrastingly large vs. minute effects observed in $\sigma_{dc}$, of externally-applied and AFM-induced $H$-fields generically relate to the spatially-uniform ($E_u$) and -modulated ($E_m$) magneto-electric fields, respectively induced by them. While $E_u$ should affect both dc and ac charge-transport (albeit differently in general), $E_m$'s effect on the dc-transport would rather nullify in its bulk-measurement, which is an average over the specimen.

Evolution of the dipolar-correlations upon cooling is obtained from the temperature dependent $\alpha(T)$ and $\beta(T)$. However, it is their physically more relevant product $\{\beta_{KWW} = \alpha\beta; \Phi(t) \sim \Phi(0)\text{Exp}[-(t/\tau)^{\beta_{KWW}}]\}$, which describes stretched-exponential time-domain behavior of the Kohlrausch-Williams-Watts (KWW) relaxations, vis-à-vis the generic exponential time-decay of independent-dipoles' ($\alpha, \beta =1$) Debyean relaxation. Moreover, it implies that a distribution of relaxation time $\{\beta_{KWW} < 1 \equiv P(\tau) \neq \delta(\tau-\tau_0)\}$ determines the dipolar dynamics [30], in the KWW correlated relaxations. The stretched-exponent parameter $\beta_{KWW}(T)$ for the $R_1$-relaxations clearly marks the antiferromagnetic (AFM) transition (~120K), whereas the abrupt drop at ~110K in that for the $R_2$-relaxations registers the emergence of WFM-nanophase in the AFM-matrix, shown in fig.6(c). Abrupt drop in $\beta_{KWW}(R_1)$ across the AFM transition implies magneto-electric enhancement of dipolar-correlations. Effective relaxation time obtained from H-N fit for $R_1$-relaxations features Vogel-Fulcher dispersion $\{\tau \sim \text{Exp}[E_a/(T-T_f)]\}$ upon cooling below 120K (~AFM-$T_N$, fig.6.(c)- top-left inset), yielding activation energy $E_a$(0T) =165.7K and freezing temperature $T_f$(0T) =50.4K. Under 6T applied field, increased $E_a$(6T)= 286.7K and reduced $T_f$(6T) =36K signify magneto-thermal activation (decorrelation) of $R_1$-relaxing dipoles. Effective relaxation time for $R_2$-relaxations (fig.6.(c)- bottom-right inset) features cluster-glass like divergence $\tau \sim (T-T_g)^{-}$

$^v$ upon cooling below 110K (~WFM-$T_W$), with $T_g$(0T) =40.K and critical exponent $v$(0T) =6.7. Under 6T field, increased $T_g$(6T) = 51K and $v$(6T) =5.7 signal magneto-consolidation of glassy $R_2$-relaxing dipoles.

From the H-N parameters at zero and 6T field, magneto-exponents M$\beta_{KWW}$(%) = 100×{$\beta_{KWW}$(6T)/$\beta_{KWW}$(0T)-1} are evaluated. Fig.6(d) shows that under the applied $H$-field $\beta_{KWW}$($R_{1,2}$) decrease below 130K (>$T_N$) itself (-ve M$\beta_{KWW}$). While the drop of $\beta_{KWW}$($R_1$) slows down and reverts back (minimum at 120K ≤ $T_N$), that of $\beta_{KWW}$($R_2$) speeds up below 115K (~$T_W$ in 6T field). These trends further corroborate our earlier inferences [17]; that in the spin-disordered state (>$T_N$), antiferrodistortive-correlations amongst both the $R_{1,2}$-type dipoles are enhanced by the applied $H$-field. Moreover, in the AFM state, the FE-correlations between $R_1$-type dipoles (hosted in the bulk-AFM phase) are field-suppressed, whereas those between $R_2$-type dipoles are further enhanced (especially below $T_W$, upon the WFM nano-phase emergence).

The temperature dependence of magneto-exponent M$\beta_{KWW}$ is found to be consistent with that of the magneto-dielectricity MD(%) =100×{$\varepsilon'$(6T)/$\varepsilon'$(0T)-1} reported earlier [18]. All-frequency -ve MD above ~$T_N$ relates to the field-enhancement of (both $R_{1\&2}$-relaxing) antiferrodistortive dipole-correlations. Positive MD at low-frequencies ($R_1$-relaxations) over 80-125K results from the field-suppressed relaxor-FE correlations of dipoles in bulk AFM-matrix. Negative MD below ~$T_N$ at higher frequencies ($R_2$-relaxations) is due to the field-consolidation of glassy dipole-correlations, to which the WFM-nano-phase owes its allegiance. Therefore, both MD and M$\beta_{KWW}$ consistently reveal the intricate dual nature of magneto-electricity in the system, which also reconciles the field-suppression of Nèel ($T_N$) and its merger with the WFM ($T_W$) temperature here [14]. Remarkably, the second-generation derived quantity $\beta_{KWW}(T)$ unequivocally registers the magnetic evolution benchmarks at $T_N$ & $T_W$. On the other hand, firstly-derived $\tau(T)$ and magneto-dielectricity MD($T$) [18] both make rather uncertain reference to the system's magneto-configuration. Contrarily, the raw dielectric data, measured at individual frequencies appears to take no cognizance of the intrinsic magnetic state. This relates to the prevailing short-ranged electrical-order, whose non-mean-field/local nature ensues dynamical heterogeneity, parameterized by timescale-distribution $P(\tau)$ [30]. Physically understood, precise magneto-electric effects manifest here in spectrally-integrated $\beta_{KWW}$ characteristic have little bearing on the spectrally-resolved quantities. The practical implication of this significance is immediate. In all systems (i.e., electrically short-range ordered), time-domain transport experiments suggest themselves as far more suitable, to directly witness the precise ME effects in the measurements. For example in $Ca_3Mn_2O_7$, the shortest times (corresponding to the fastest relaxation peak-frequency at ~300 kHz, say) work out as $\tau \sim O(10^{-6}$ sec). A time-resolved transient-conductivity measurement lasting at least ~10 μ-seconds, in response to a step-up/down *E*-field excitation is rather viable using fast instrumentation. Temporal characterization of the measured $\sigma_T(t)$ isotherms in such experiments is thus expected to directly reveal the magneto-electric imprints of the system.

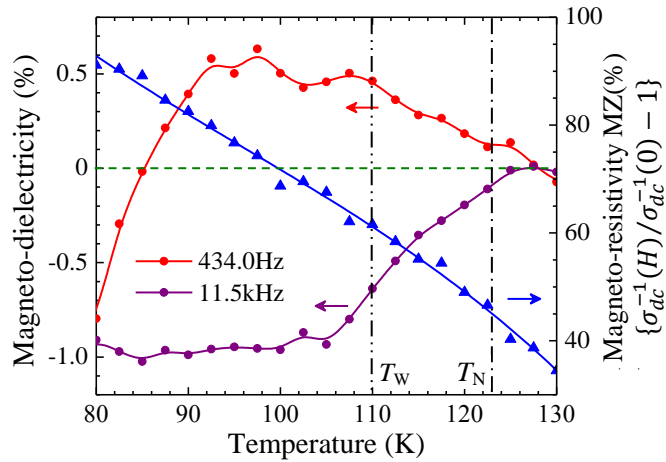

Figure 7. Temperature dependence of MD at 434 Hz and 11.5 kHz (left pane), featuring behaviors uncoupled with the MR (right pane).

From the Jonscher-fit $\sigma_{dc}$, we define the dc 'magneto-resistivity' as MZ =$\{\sigma_{dc}^{-1}(H)/\sigma_{dc}^{-1}(0) - 1\}$; expected to reflect the behaviour of the usual magneto-resistance MR. MZ is seen to increase upon cooling, and is anomalously incognizant of magnetic state evolution (fig.7: right pane); consistent with the nearly-vertical steps in $n(\sigma_{dc})$, fig.6(b). Figure 7 also shows MD($T$) [18] at 434 Hz and 11.5 kHz, mainly contributed to by the dipolar organizations undergoing $R_1$- and $R_2$-relaxations respectively. MD($T$) generically illustrate magneto-electric evolution of the system, and lack an algebraic co-variation with the feature-free MZ($T$), which registers an order of magnitude larger and essentially monotonic change ($\Delta$MZ =55%) over the ME-relevant temperature range 80-130K. This evidences that the (ac-) magneto-electricity in the system is exclusive of the (dc-) magneto-resistivity, as e.g. reported previously for $La_{0.53}Ca_{0.47}MnO_3$, via the latter's direct measurement [31].

Comparatively low MD magnitude (~1% max.) obtained vis-à-vis rather dramatic changes in the relevant sub-structural attributes (fig.3(c)) are understood as follows. In the $n$ =2 Ruddlesden–Popper (RP) compounds, because of mutually 'antiferroelectric' displacements of A-cations in seven AO layers within a unit cell, there results only a small net dipole moment from the 'uncancelled' AO-layer [5]. For $Ca_3Mn_2O_7$ system, lack of long range electrical order plus cancellation of A-cations' vector-displacements therefore account for the low MD. Note that for the most part, both |MD| and MZ increase upon cooling. Extending the analysis by Catalan [32]-- carried out for negative magneto-resistance systems, to the present case; |MD| would be expected to anti-regress with the +ve MZ, if magneto-capacitance was manifest without any magneto-electric coupling. Therefore, the +ve regression observed here between |MD| and MZ again discounts the magneto-resistive origins of genuine magneto-electricity in the system.

## CONCLUSIONS

We have established prominent spin-lattice coupling in $Ca_3Mn_2O_7$ by analyzing its Synchrotron-based structural data from 300K down to 37K, in close temperature steps. Appreciable decrease observed in the particulate bond-angle corroborates the emergence of a WFM-nanophase. Upon cooling across the AFM transition, slower decrease of the unit cell volume along with dramatic rise of the displacement vector defining net dipole moment, are observed. These structural changes signify large spin-lattice coupling and its enhancement upon the formation of interacting-dipoles nano-regions. Spectral analysis of dipole relaxations corroborates the dual (±ve) nature of magneto-dielectricity in the system, besides marking its magnetic evolution, consistent with the structural signatures obtained. Low values of magneto-dielectricity observed here vis-à-vis large structural changes concurrent to spin-ordering, is reconciled by a major antiferroelectric-cancellation of the intra unit-cell dipole moments. Finally, magneto-electricity as the convoluted manifestation of its coupled spin, lattice, and charge degrees of freedom is confirmed to be genuine, independent of its magneto-resistance.

As outlook, the issue of rather small magneto-electricity, despite a strong spin-lattice coupling, could be addressed by partly replacing the AO-layers by A'O-layers within the unit cell; thus realizing the $(ABO_3)_2A'O$ structure. Further substitution of A'-cation may reduce the antiferroelectric-cancellation of intra-cell layer-polarizations. Another approach to introduce strain in the system is dimensional reduction— e.g., nanostructuring, thin-film, and nano-fiber synthesis. Tuning down the antiferrodistortive energetics in the system, to enhance the unit cell polarization, therefore seems the key to robust ferroelectricity prospects, alongwith its realization at ambient temperatures, for functional magneto-electricity. Our study also naturally suggests the transient time-domain transport investigations of electrically short-range ordered systems, for directly witnessing their magneto-electric character, often subtly-hidden and easily-missable in their frequency-resolved dielectric spectroscopy.

## ACKNOWLEDGMENTS

We deeply appreciate M.N. Singh (BL-12, RRCAT) for helping with the SXRD measurements. We extend our sincere thanks to Suresh Bhardwaj (UGC-DAE CSR) for help with the dielectric measurements.